\documentclass[]{article}

\usepackage{mathptmx}       % selects Times Roman as basic font
\usepackage{helvet}         % selects Helvetica as sans-serif font
\usepackage{courier}        % selects Courier as typewriter font
\usepackage{type1cm}        % activate if the above 3 fonts are
\usepackage{makeidx}         % allows index generation
\usepackage{graphicx}        % standard LaTeX graphics tool
\usepackage{multicol}        % used for the two-column index
\usepackage[bottom]{footmisc}% places footnotes at page bottom
\usepackage{url}

\begin{document}

\begin{flushleft}
{\Large
\textbf{Towards a comparative science of cities: using mobile traffic records in New York, London and Hong Kong\\[10pt]}
}
% Author names, affiliations and corresponding author email.
S\'ebastian Grauwin$^{1,\ast}$, 
Stanislav Sobolevsky$^{1}$,
Simon Moritz$^{2}$,
Istv\'an G\'odor$^{3}$,
Carlo Ratti$^{1}$
\\[10pt]
\bf{1} Senseable City Laboratory, Massachusetts Institute of Technology, Cambridge, Massachusetts, United States of America
\\
\bf{2}  Ericsson Research, Sweden
\\
\bf{3} Ericsson Research, Hungary
\\
$\ast$ E-mail: sgrauwin@mit.edu
\end{flushleft}

%\author{Sebastian Grauwin, Stanislav Sobolevsky, Simon Moritz, Istv\'an G\'odor and Carlo Ratti}
\date{}

\abstract{This chapter examines the possibility to analyze and compare human activities in an urban environment based on the detection of mobile phone usage patterns. Thanks to an unprecedented collection of counter data recording the number of calls, SMS, and data transfers resolved both in time and space, we confirm the connection between temporal activity profile and land usage in three global cities: New York, London and Hong Kong. 
By comparing whole cities typical patterns, we provide insights on how cultural, technological and economical factors shape human dynamics.
At a more local scale, we use clustering analysis to identify locations with similar patterns within a city.  Our research reveals a universal structure of cities, with core financial centers all sharing similar activity patterns and commercial or residential areas with more city-specific patterns. These findings hint that as the economy becomes more global, common patterns emerge in business areas of different cities across the globe, while the impact of local conditions still remains recognizable on the level of routine people activity.}

%----------------------------------------------------------------------------------------------------------------------------------------------------------------------------
%----------------------------------------------------------------------------------------------------------------------------------------------------------------------------

\section{Introduction}
\label{section::intro}
As digital technologies are becoming more and more widespread, big data created by recording the digital traces left behind human activities become a powerful mean to study various aspects of human behavior. Many of these aspects can be described with telecommunications data which nowadays become global. The exploration of these data provides new perspectives, revealing characteristic usages and regular dynamic patterns at both the individual and collective scale.
At the same time, the increasing urbanization of the world's population is deeply affecting urban environments, and it is crucial to develop theoretical frameworks as well as real-time monitoring systems to understand how the individual dynamics shape the structure of our cities in order to make better planning decisions. 

In the past years, several studies have shown that it was possible to use telecommunication data to get a fresh view at the spatio-temporal dynamics within a city.
In a now-famous paper, Eagle and Pentland \cite{eagle2006reality} showed that it was possible to decompose mobile phone activity patterns of university students into regular daily routines, and that these routines were linked to each student's major and also to employment levels. Building upon this work, Gonz\'alez et al. \cite{gonzalez2008mobilitypatterns} studied the trajectory of 100,000 anonymized mobile phone users to reveal statistical regularities in human trajectories. This paper, along with other seminal work \cite{candia2008uncovering, song2010limits} has since generated a research field dealing with human mobility as understood from digital traces \cite{kang2013exploring}.

In parallel, focusing on records aggregated on spatial locations rather than on individuals, new approaches have been initiated to describe urban landscape based on mobile phone usage patterns \cite{jacobs2012linking, loibl2012mobile,ratti2006mobile, reades2007cellular, reades2009eigenplaces, sun2011exploring}, to explore the issue of regional delineation \cite{amini2014impact, kung2013exploring, ratti2010redrawing,sobolevsky2013delineating}, to estimate population density \cite{girardin2009towards, kang2012towards,  rubio2013adaptive,vieira2010characterizing} or to identify social group and social events \cite{traag2011social}. In particular, by measuring mobile phone data on a 500m by 500m `pixel' grid in Rome (Italy), Reades et al. \cite{reades2009eigenplaces} especially used a variant of principal component analysis to cluster these pixels into regions with similar patterns of usage, and made a qualitative link between these patterns and the number of businesses on the corresponding areas. 

This last paper is an example of a line of research dealing with the identification of specific land use type \cite{caceres2012,calabrese2010eigenplaces}. Other papers have focused on methods to build classification of several land use type based either on (voice calls or SMS) mobile phone patterns \cite{andrienko2013multi,becker2011tale, pei2013new,soto2011automated, toole2012inferring}, taxi trip data \cite{liu2012urban} or Twitter data \cite{frias2012characterizing}. These studies used different types of methods, from simple clustering to advanced neural network models. A common feature of these papers is that they are limited to the study of a single spatial entity (in general a city) that they study through one type of digital data.

This statement raises some questions: is the behavior detected on one type of mobile phone activity independent of other type, i.e. is it the same to look at calls, SMS or even data transfers? How does the results compare between multiple cities? What are the signatures of the mobile network usage in major US, European or Asian cities and how do they compare? 

This chapter takes advantage of an unprecedented multi-modal collection of counter data recording the number of calls, SMS, request and data transfer resolved both in time and space in three cities - New York, London and Hong Kong - to investigate such questions. 
After presenting the data in section \ref{section::matmeth}, section \ref{section::spatial} will investigate the spatial repartition of activity in these three cities.
Then in section \ref{section::temporal}, we will focus on people's behavior by investigating the dynamics of the activities on both a local and city scale. 
In section  \ref{section::cluster}, we will show how we can use a clustering algorithm to automatically detect and classify such patterns either within one city or across all of them.
Finally, we will discuss in section  \ref{section::discussion} how our work can help us understand the changing nature of modern cities and especially what common features can be captured in the patterns of human behavior: in these global cities, what is the respective influence of city-specific and global factors on human life?

%----------------------------------------------------------------------------------------------------------------------------------------------------------------------------
%----------------------------------------------------------------------------------------------------------------------------------------------------------------------------

\section{Materials and Methods}
\label{section::matmeth}
\subsection{Geographical background}
\label{section::geobackground}
\begin{figure}[h!]
   \begin{center}
	\includegraphics[width=1\textwidth]{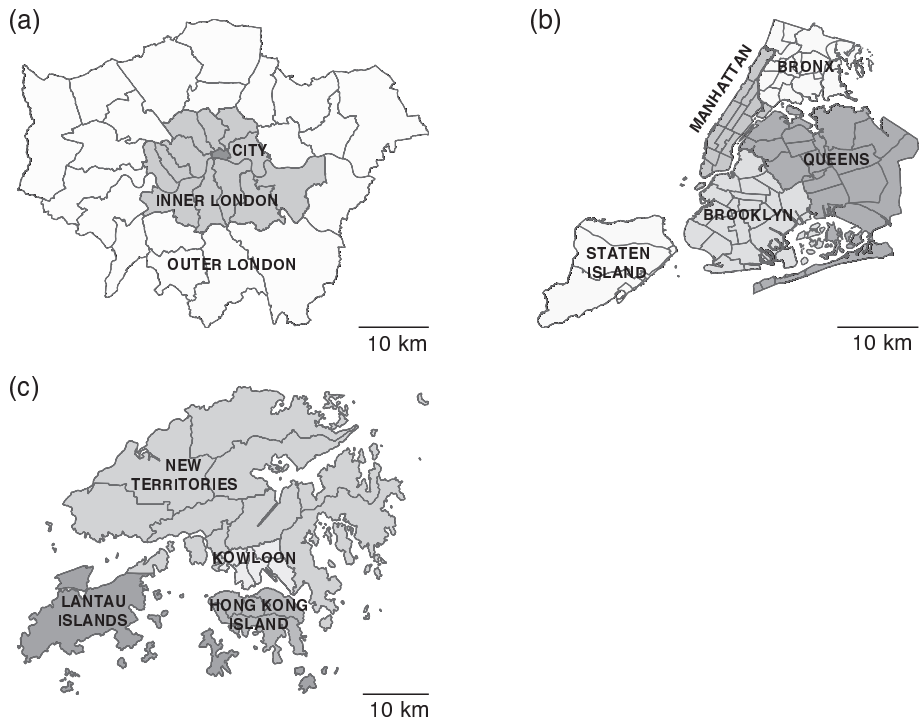}     
   \end{center}
   \caption{{\small {\bf Administrative maps of} {\bf (a)} Greater London {\bf (b)} New York and {\bf (c)} Hong Kong.\label{fig1}}}
\end{figure}

Figure \ref{fig1} shows a map of the three cities studied in this chapter: New York, London and Hong Kong. 
Greater London is divided into 33 `district boroughs', the central ones being referred to as {\it Inner London}, while the peripheral ones are referred to as {\it Outer London}. The historic heart of London, the {\it City of London}, is a major business and financial centre, where are located many banking and insurance institutions headquarters. While the City has a low resident population (around $7000$), over 300,000 people commute and work there every day, mainly in the financial service sector.
New York is divided in 59 `community districts', gathered into five boroughs. The boroughs of Queens, Brooklyn, Staten Islands and Bronx are manly residential. Manhattan is a major financial and decision-making centre (the UN headquarters, Times Square and the Empire State Building are located there). It has also one of the highest population density in the world, with around 27,000 residents per square kilometers. 
Finally, Hong Kong territory consists of four regions split into 18 districts. Due to the mountainous nature of its land, less than 25\% of Hong Kong's territory is urbanized: the urban development concentrates on Kowloon peninsula, the northern edge of Hong Kong Island and a few settlements throughout the New Territories. 

Overall, London, New York and Hong Kong are comparable in terms of size and population density (see Table \ref{table1}). They are also all Alpha cities according the GaWC nomenclature of world cities \cite{beaverstock1999roster} which ranks cities based on their connections with others in domains such as business, finance, law, media, art, fashion, research, technology, education, tourism and entertainment. As Alpha cities, they are more integrated within the global economy than any other city. Despite their apparent differences (in terms of history, culture or weather conditions to name only the obvious ones) and geographical distances, one can expect emerging similarities between these cities due to globalization. They are therefore perfect candidates for tracking universal communication patterns.

\begin{table}[h!]
\caption{Background information on the scale of the studied cities (source: wikipedia, 2013). \label{table1}}
\vspace{1mm}
\begin{tabular}{ lccc}
\hline\noalign{\smallskip}
City & Area ($km^2$) & Population & Density ($pop/km^2$)\\
\hline\noalign{\smallskip}
Greater London & $1572$ & $\sim 8,200,000$ & $5206$\\
New York & $1214$ & $\sim 8,350,000$ & $6865 $\\
Hong Kong & $1104$ & $\sim 7,000,000$ & $6405 $\\
\hline\noalign{\smallskip}
\end{tabular}
\end{table}

\subsection{Data gathering and pre-processing}
\label{section::data}
Our analysis is based on aggregate 3G mobile traffic data (including all kind of devices like phones, tablets, etc.) supplied by several operators, and corresponding to a statistically significant part of the total 3G mobile traffic of the covered areas, corresponding to several millions subscribers in all studied cities (precise penetration rates can not be given for confidentiality reasons). Data were collected between April 1st and July 7th 2013 at fifteen minute intervals across the three cities at cell level. 
While the whole city of New York is covered, the dataset of Greater London is mostly concentrated on the Inner London districts. The Hong Kong measurements cover urban zones, while no data is available for the unpopulated mountainous parts.  

The three months of data consist in counter data recording the numbers of Calls, SMS and Requests (for data communication initiated either by the users or some applications running in the background in their mobile devices) as well as the amount of data uploaded and downloaded by subscribers (measured in Bytes and thereafter denoted by `UL Data' and `DL Data'). 
The provided data was aggregated at the cell level by the data providers and therefore did not reveal any individual user information. Before receiving the data, the actual numbers were obfuscated by using a secret scaling factor, such that we have only access to normalized amounts of each type of counter data.

In addition to mobile phone data, we gathered various shapefiles, census data and land use data from open access sources\footnote{such as \url{http://data.london.gov.uk/} for London, \url{https://nycopendata.socrata.com/} for New York or \url{http://www.census2011.gov.hk/} for Hong Kong}. We thus obtained land use data with nature and number vary across the different cities (9 categories for London, 6 for New york, 24 for Hong Kong). To better compare the results obtained in our three cities, we converted these original categories into seven land use type that best match them: High Density Residential, Low Density Residential, Business \&  Commercial, Mixed (Residential \& Commercial), Infrastructures, Parks, Other. Details of the procedure are available on request.

The location of the cells recording mobile phone activity is given as longitude / latitude pairs - the service area of a cell having a typical radius varying from around 100m (in dense area) to several kilometers (in rural zones), while the census and land use data are provided in polygonal zones corresponding to administrative divisions in the cities. In order to  study the mobile phone usage patterns in the different cities and their relationships to census and land use data, we chose to transform the spatial representation of the different datasets by projecting them on uniform lattice grids of 500m by 500m `pixels'.

To reduce the bias induced by the attribution of the activity within a cell's service area to a single pixel location, we used a smoothing procedure: we defined the activity on one pixel as the mean of activities of all cells within a 1500m by 1500m square centered on the pixel center. Census data were similarly projected on the grid by interpolating demographic data and most significant land use on each grid cell.
The length of 500m was chosen after testing different grid sizes. It proved to be coarse enough to reduce noise level, and detailed enough to explore spatial patterns of activities within the cities. At the end of this procedure, the mobile phone traffic data was projected on around $2700$ pixels in Greater London and around $3000$ pixels in New York and Hong Kong.

%----------------------------------------------------------------------------------------------------------------------------------------------------------------------------
%----------------------------------------------------------------------------------------------------------------------------------------------------------------------------

\section{Spatial repartition of activity}
\label{section::spatial}
A simple question that can be addressed with mobile communication data is the tracking of where people are or where they go. 

\begin{figure}[h!]
   \begin{center}
    \includegraphics[width=1\textwidth,trim = 0cm 0cm 0cm 0cm, clip]{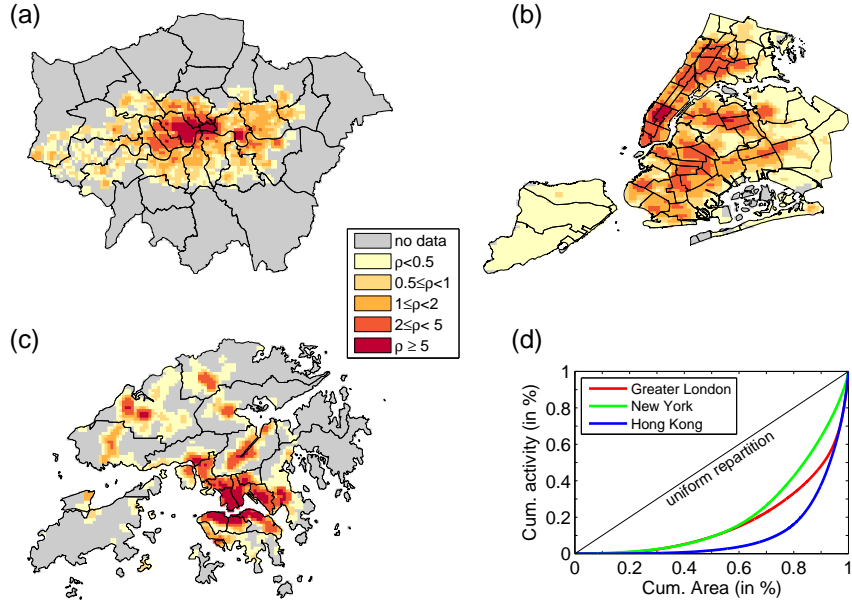} 
   \end{center}
   \caption{{\small {\bf Spatial repartition of activity}. {\bf (a), (b), (c)} show the values $\rho$ of normalized total request activity aggregated over the three months of data on the pixel grids of resp. Greater London, New York and Hong Kong. {\bf (d)} Lorentz curves showing the variation of cumulative total request activity with the cumulative area of coverage, the pixels being ranked by increasing total request activity. \label{fig2}}}
\end{figure}

A first way to investigate this question is displayed on Figure \ref{fig2}(a-c), which shows the spatial repartition of total request activity as recorded in our datasets. 
We chose to focus on the request activity since it passively tracks people: even if you don't make any call or send any SMS, your mobile device will still produce network background traffic such as social network synchronization, weather updates, news feeds, etc. 

The colors used on the maps of Figure \ref{fig2}(a-c) emphasize the inhomogeneities of activity repartition, by showing the share $\rho$ of activity on each 500m by 500m pixel normalized by the total activity recorded. A value of $\rho=1$ hence corresponds to a pixel with average total activity, $\rho> 1$ to a pixel with higher than average activity and $\rho < 1$ to a pixel with lower than average activity. 

The spatial repartition of $\rho$ in the three cities shows a center / periphery dichotomy. 
In Greater London, we observe a concentric organization, with a very strong activity level in the City of London, and decreasing levels as one move away from the city center.
In New York, the organization is polycentric, with one center in the middle part of Manhattan, and another one in Queens. 
Finally, in Hong Kong we observe one big center of activity divided up between Kowloon and the northern part of Hong Kong Island and secondary centers in the newly developed zones of the New Territories. Low activity zones correspond to the limits with the mountainous areas.

\begin{table}[h!]
\caption{{\small {\bf Gini coefficients}} For each activity type $\lambda$, the Gini coefficient $G^{\lambda} \in [0, 1]$ measures the inhomogeneity of the spatial repartition of the three-months aggregated activity. The higher the coefficient, the more unequal the spatial repartition is.  \label{table::ginis}}
\begin{tabular}{lrrrrr}
\hline\noalign{\smallskip}
City & $G^{DL\,Data}$ & $G^{UL\,Data}$ & $G^{Request}$ & $G^{Calls}$ & $G^{SMS}$\\
\hline\noalign{\smallskip}
Greater London & 0.608 & 0.640 & 0.649 & 0.618 & 0.606 \\
New York & 0.546 & 0.555 & 0.576 & 0.549 & 0.523 \\
Hong Kong & 0.768 & 0.765 &  0.784 & 0.802 & 0.781\\ 
\hline\noalign{\smallskip}
\end{tabular}
\end{table}

Lorentz curves depicting the variation of cumulative total request activities with the cumulative areas of coverage are presented on Figure \ref{fig2}(d). These curves, typically used in economy and ecology to describe inequality in wealth or size \cite{lorenz1905methods}, describe here unequal repartition of request activity in space. 
Hong Kong is obviously the most inhomogeneous city (with less than $2\%$ of request activity in the $50\%$ less active pixels and  $64\%$ of activity concentrated in the top $10\%$ most active pixels), followed by London (around $9\%$ of activity in the $50\%$ less active pixels and $49\%$ of activity in the top $10\%$ most active pixels) and New York (around $9\%$ of activity in the $50\%$ less active pixels and $38\%$ of activity in the top $10\%$ most active pixels).

% --- 
A commonly used quantitative measure of inequality, the Gini coefficient, can be defined from a Lorentz graph as the area between the bisector line (corresponding to a uniform repartition) and the Lorentz curve, normalized by the area between the the bisector line and the x axis (corresponding to the most inhomogeneous case where all activity is concentrated on one pixel) \cite{gini1912}. 
The Gini coefficients of the request Lorentz curve of Figure \ref{fig2}(d) as well as those corresponding to other types of activity are reported in Table \ref{table::ginis}. 
Interestingly, the Gini coefficient depends only slightly on the activity type and strongly on the city. The measure of spatial inhomogeneity could thus be done on any type of activity. Again, we find here that the most inhomogeneous city is Hong Kong, followed by London and then New York.

Although this first analysis suffers from some limitations - such as the mismatch between the area covered by our dataset in London and the area within the official boundaries of Greater London - it already provide good insights on the way people interact with their cities. Maps of mobile phone activities could steadily become a complementary tool to more classical maps of population or employment densities obtained through extensive surveys, and help urban planners make decisions based on accurate population repartition.

%----------------------------------------------------------------------------------------------------------------------------------------------------------------------------
%----------------------------------------------------------------------------------------------------------------------------------------------------------------------------
\section{Exploring temporal patterns}
\label{section::temporal}
%Let's now explore the dynamics of the mobile phone activity.

\subsection{Typical week signature} 
\label{section::typweekdef}
To minimize the impact of special events on the datasets, we followed the procedure presented in \cite{reades2007cellular} to extract average `typical week' timelines for each pixel at each 15 min interval, using the three-months period. 
The values of the typical weeks for a given 15 min time interval were calculated as the average of the same intervals from the whole measurement. For example, the typical number of calls  for 12:00 to 12:15 on the typical Monday was taken as the average number of calls from 12:00 to 12:15 on every Mondays available in the dataset.
Civic holidays, considered as special events, were excluded from the computation to avoid the introduction of unnecessary noise. 
We use these mathematical notations:
\begin{itemize}
\item $A^{\lambda}_i(t)$, to measure activity of type $\lambda$ (number of Calls, SMS or Request, volume of Data upload or download) within a given pixel $i$ at time $t\in[1,672]$ (since the measurements are taken every 15min, one week comprises $7\times96=672$ time intervals).
\item $A^{\lambda}_{city}(t)$, to measure activity of type $\lambda$ (number of Calls, SMS or Request, volume of Data upload or download) within a given $city$ at time $t\in[1,672]$.
\end{itemize}

% In order to compare the dynamic patterns in two different locations, it is useful to normalize these values to focus on the behavior of people rather than on their sheer number. 
In order to better compare the relative dynamic patterns across the cities' locations (e.g recognizing locations with similar patterns up to multiplicative constant due to higher active population density), it is useful to normalize these values by the typical amplitude of activities on each pixel. We thus define the Signature $S^{\lambda}_i$ of activity type $\lambda$ on location $i$ thanks to a mean normalization: 
\begin{equation}
S^{\lambda}_i(t) = A^{\lambda}_i(t) / \langle A^{\lambda}_i\rangle_t ,
\end{equation}
where $\langle ...\rangle_t$ denotes an average over the 672 individual 15-min time intervals of a typical week. Similarly, the Signature $S^{\lambda}_{city}$ of activity type $\lambda$ at the city scale is given by: 
\begin{equation}
S^{\lambda}_{city}(t) = A^{\lambda}_{city}(t) / \langle A^{\lambda}_{city}\rangle_t.
\end{equation}

\begin{figure}[h!]
   \begin{center}
      \includegraphics[width=1\textwidth,trim = 1cm 0cm 1cm 0cm, clip]{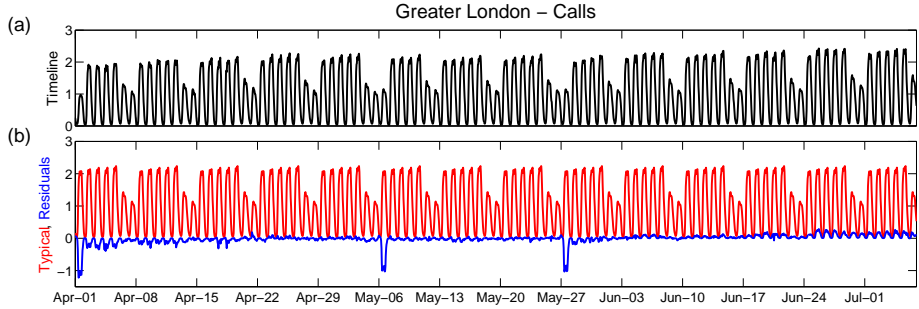}   
   \end{center}
   \caption{{\small {\bf Decomposition of Greater London `Calls' Timeline.} {\bf (a)} The mean normalized timeline can be decomposed into {\bf (b)} a repeating weekly pattern (the city Calls signature $S_{London}^{Calls}$, in red) and a residual part (in blue). \label{fig3}}}
\end{figure}

As an illustration of the computation of Signatures, Figure \ref{fig3}a displays the mean-normalized Calls timeline in Greater London over the three month observation period. This timeline shows daily variations - with peaks of activity during the days and drops of activity at nights -, and weekly variations, with daily peaks significantly lower during weekends than during workdays. Figure \ref{fig3}b then shows how this timeline can be decomposed into a repeating typical week pattern and a residual part.
The residual part accounts for special events, such as the occurrence of civic holidays (notice the lower amount of Calls on April 1st, May 6th and May 27, respectively Easter, May Bank and Spring Bank holidays in London), and general trends. For example, we observe a slight overall increase in Calls in late Spring / early Summer which may be due to the arrival of tourists in London at this time of the year.

Overall, the typical week signature captures the main temporal patterns within the cities, by reducing noise and getting rid of long term trends. Incidentally, it can provide a good predictive baseline of expected mobile device usages. In the example presented in Figure \ref{fig3}, the average absolute ratio between residuals and timelines is approximatively equal to $8\%$, but is typically included between $5$ and $10\%$ for the different activity type either in whole cities or at each pixel level. 
Based on this timeline, an operator could detect irregular operations, as also anticipate upcoming special events.

%----------------------------------------------------------------------------------------------------------------------------------------------------------------------------
\subsection{Comparing Cities' Signatures}
\label{section::typweekcities}
\begin{figure}[h!]
   \begin{center}
      \includegraphics[width=1\textwidth,trim = 0cm 2cm 1cm 0cm, clip]{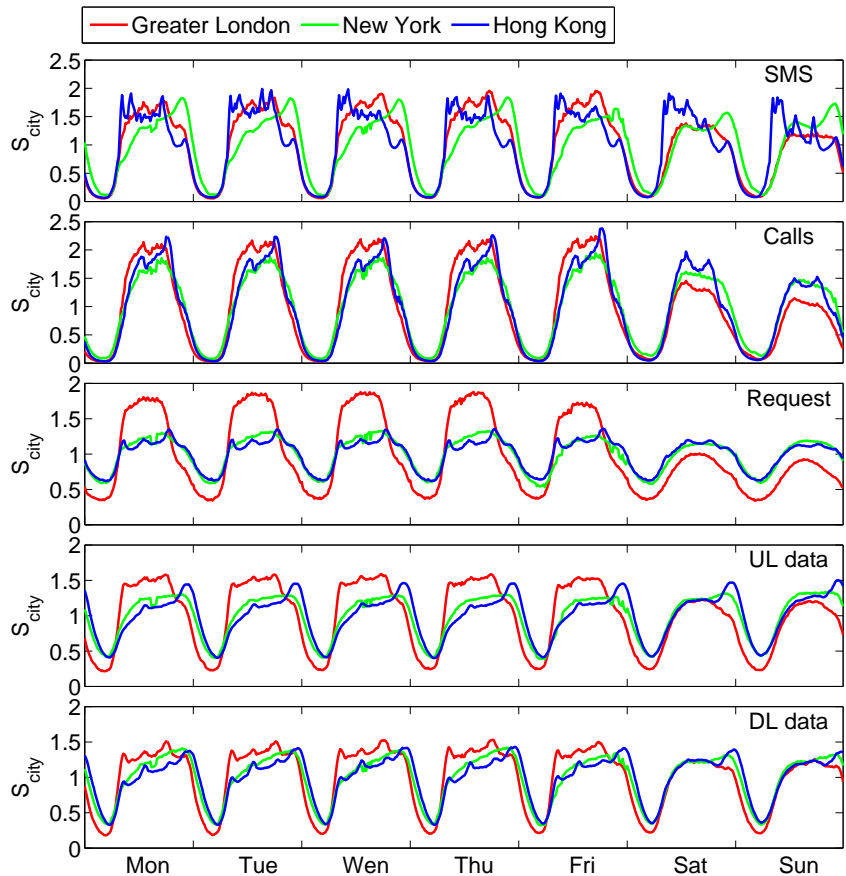}
   \end{center}
   \caption{{\small {\bf Cities signatures}, showing the normalized typical week patterns for the different components of activity at the city scale. \label{fig4}}}
\end{figure}

Let us first explore the temporal patterns at the macro, city-wide scale. 
The city signatures of the different activity types are displayed on Fig. \ref{fig4} which emphasizes their similarities and differences.

All three studied cities display a broadly comparable rhythm, common to all components of activity. Mobile activity rapidly ramps-up in the morning between 6 and 10 AM, followed by rather steady activity levels within the day, and a slower decrease of activity at night between 9 PM and 2 AM. The same pattern appears on workdays and with somewhat slower amplitude on weekends. On workdays, we can also observe small peaks of activity at commuting hours and at lunch times. It thus seems reasonable to associate this common rhythm to a simple daily cycle, corresponding to people waking up, going to work, having lunch and then heading back home in the evening. 

Let us now turn to the differences appearing on Figure \ref{fig4} when comparing the cities.
\begin{itemize}
\item The most obvious difference can be seen on the curves displaying the request signature. The London request signature presents a daily cycle with higher variations than the New York and Hong Kong ones, and its relative drop of activity in the weekend compared to workdays is also more important. The cost of mobile data plans being higher in London than in other parts of the world, our educated guess is that Londoners use (cheaper) wifi whenever available to connect their mobile devices. Since these connections are not recorded in our dataset, one can expect a specific negative bias in the data when people typically switch from the operator mobile network to wifi network: in the evening and on the weekend, when they are at home.
\item The early evening drops of UL and DL Data signatures in London can be explained in the same way. Londoners do not necessarily go to bed or stop to use their mobile devices earlier, but more probably use their home wifi connections more often.
%\item The drop of activity in the weekend compared to workdays is more significant in London than in Hong Kong or New York, whatever the activity type. This could be explained by taking into account the residential capacity of the cities' business areas. In New York and Hong Kong the business areas are also where people work while in London the City of London generating a significant part of mobile phone activity (as can be seen on Figure \ref{fig2}) is purely a financial place.
\item In both New York and Honk Kong, the daily shape of all activity types is slightly asymmetric, the maximum activity being reached in the evenings. Surprisingly, there is an exception with the SMS activity in Hong Kong that appears to drop earlier than the other activities in the evenings. Rather than a disinterest from Hong Kongers towards texting in the evening, our guess is that the Hong Kong SMS signature reveals a great specific interest towards texting during the day. Text messaging is indeed known to be particularly popular in Asia, where companies use text messages to confirm deliveries and provide alerts, updates or infotainment. 
\item We also observe peaks of SMS activity on evenings in New York. These peaks could be explained by the important use of SMS for media voting (e.g. on TV show polls) in the US like America's got talent or X-factor.
\end{itemize} 

At  first glance, New York and Hong Kong, while located almost at opposite places of the globe and having different cultural background, may surprisingly appear to have more similar signatures than New York and London which share a common cultural and linguistic background. However we have seen that these signatures are shaped by many different (technological, economical or cultural) factors and that their interpretation must reflect the multiple influences on people's behaviors. Space is also an important factor to take into account since people do not behave in the same way depending of where they are. 

 %----------------------------------------------------------------------------------------------------------------------------------------------------------------------------
\subsection{Comparing Local Signatures}
\label{section::typweeklocal}
In this section, we go further in our exploration of what we can learn about people's bevahiors based on their communication patterns. After comparing cities' signatures, we will now compare local signatures from different locations within a same city. 

\begin{figure}[h!]
   \begin{center}
      \includegraphics[width=1\textwidth,trim = 0cm 0cm 0cm 0cm, clip]{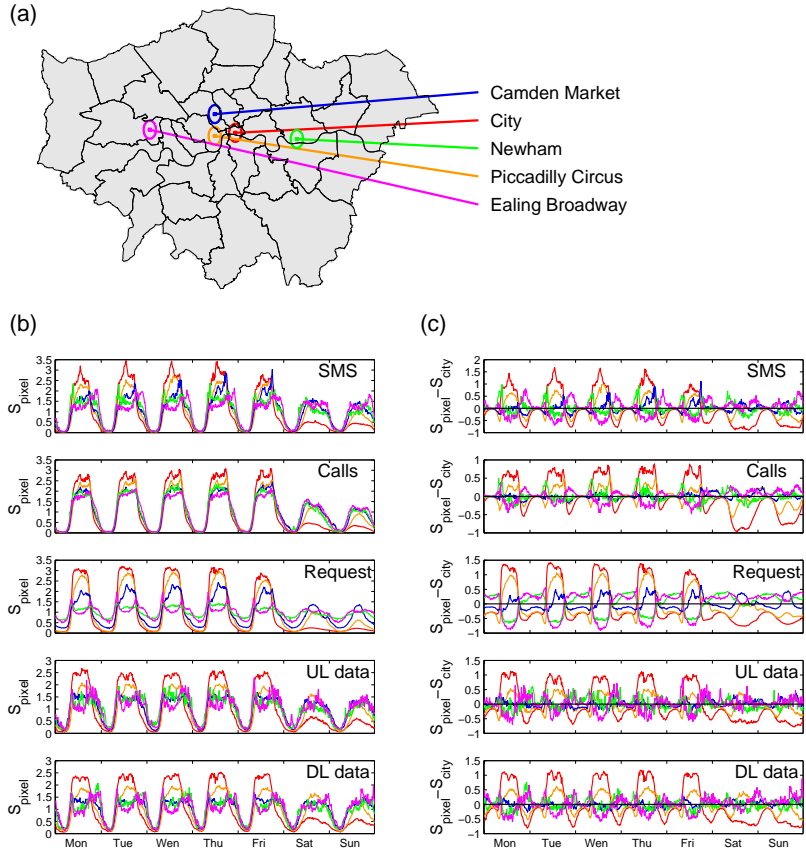}
   \end{center}
    \caption{{\small {\bf Local signatures} {\bf (a)} We selected five 500m by 500m grid pixels in specific locations of Greater London. {\bf (b)} The first series of plots show these locations' signatures in the different components of activity. {\bf (c)} The second series of plots shows the differences between local and whole city signatures. Colors in the plots match those on the map indicating the locations of the selected pixels.\label{fig5}}}
\end{figure}

As shown on Figure \ref{fig5}(a), we selected five different 500m by 500m pixels corresponding to specific locations in London: one pixel in the City of London (the financial centre of London), one centered on Piccadilly Circus (a public space in Westminster close to major shopping, entertainment and touristic area), one on Camdem Market (a popular market place where crafts, clothing's and fast-food are sold, especially on weekends), one on Newham (a residential area) and one on Ealing Broadway (a travel hub, part of the National Rail and London Underground networks).

Figure \ref{fig5}(b) shows the signatures of all activity types on each of these five locations, while Figure \ref{fig5}(c) highlights the specificities of each location's signature by displaying their deviation from the whole Greater London's signature.

Similarly to the city level, these five sample signatures display a comparable day~/~night cycle,  but also present specific characteristics revealing the nature of the corresponding locations.
\begin{itemize}
\item The signatures of the pixel within the City of London show patterns typical of a business area: high amount of activity during working hours and very low activity in the evening from Monday to Friday and huge weekday-to-weekend activity ratios. These signatures also display sharp transitions from low to high activity level at the beginning of working hours and from high to low level at the end of working hours.
\item The Piccadilly Circus signatures also show a large amount of activity during working days, but also show significant activity during the weekend. The morning transitions from no activity to some activity is rather smooth (notice in particular the dips on Figure \ref{fig5}(c) revealing that this area is less active than other parts of the city in early morning). All these characteristics go in line with the commercial and touristic nature of the place. 
\item The Camden Market signatures have average patterns during working hours and are typically characterized by a peak of SMS activity during the workdays evening (consistent with the recreational nature of this location where people may gather to share a drink) and high level of request activity around lunch time and in the early afternoon in the weekend (in line with the popularity of the markets).
\item The signatures of the Newham pixel are specifically characterized by low weekday-to-weekend activity ratios (suggesting a constant population rate over the week) and specific non-zero request activity at night (the automatic update of the mobile devices revealing that people are sleeping at those locations). This characteristics straightforwardly reveal the residential nature of this location.
% schema inverse request
\item Finally, the signatures in Ealing Broadway are rather comparable to Newham's one - revealing the residential nature of the area - but also show specific peaks of request activity during morning and evening commuting hours, in line with the commuter hub property of the Underground station.
\end{itemize}

These observations show a correspondence between mobile traffic signatures and the nature of the places studied, suggesting that we could infer the nature of each area based on its signatures. 
As a side perspective, the identification of the central business district and the residential areas can offer insights on the nature of commuting flows. Urban planners are already aware of the spatial relationship between business district and residential areas, but the visualization of such properties on such a precise spatio-temporal scale has been possible only for a few years thanks to the advancement of digital data gathering.

%----------------------------------------------------------------------------------------------------------------------------------------------------------------------------
%----------------------------------------------------------------------------------------------------------------------------------------------------------------------------

\section{Cluster Analysis}
\label{section::cluster}
\subsection{Principles}
\label{section::clusterdef}
In the previous section, we focused on single pixels, and we have shown how their signatures could reveal human dynamics' features at the local level. 
In this section, we investigate the question of whether we can use mobile devices traffic data to detect large areas with homogeneous properties. Our goal is to group local pixels according to the similarity of their signatures and use these groups to map the urban spatiotemporal structure of the cities.

Among the many different clustering techniques to extract clusters of pixels with similar signatures, we chose a K-means approach, used in many previous studies \cite{andrienko2013multi, pei2013new,reades2007cellular}. This approach ensures that each pixel of a cluster has a signature as much like the one one the other members of the clusters and as different as possible from the signature of the pixels in the other clusters. 
Starting from the signatures $ \{ S_i^{\lambda} \}$ of the pixels, the signature of a cluster $C$ is defined as the average signature of the pixels within that cluster:
\begin{equation}
S_{C}^{\lambda}=\langle S_i^{\lambda} \rangle_{i\in C}
\end{equation}

Combining all activity types, the K-means algorithm aims at minimizing the quantity $E_K$ measuring the total distance between the locations' signatures and their cluster's signature:
\begin{equation}
E_K=\sum_{k=1}^K \sum_{i \in C_k}  dist(i, C_k),
\end{equation}
where the distance $dist(i,C)$ between a pixel $i$ and a cluster $C$ is defined as
\begin{equation}
dist(i, C)=\sum_{\lambda} \sum_{t} \left(S_i^{\lambda}(t) - S_{C}^{\lambda}(t)\right)^2,
\end{equation}
and $K$ is a pre-imposed number of clusters. 
This simple quantity does not take into account the temporal structure of the signatures (the order of the different time intervals do not matter), but in the following, it will prove to deliver consistent results. Each pixel is characterized here by a 3360-dimensional feature vector (5 signatures of different type, each being valued on $672$ time intervals). 

\begin{figure}[h!]
   \begin{center}
      \includegraphics[width=0.99\textwidth,trim = 0cm 0cm 0cm 0cm, clip]{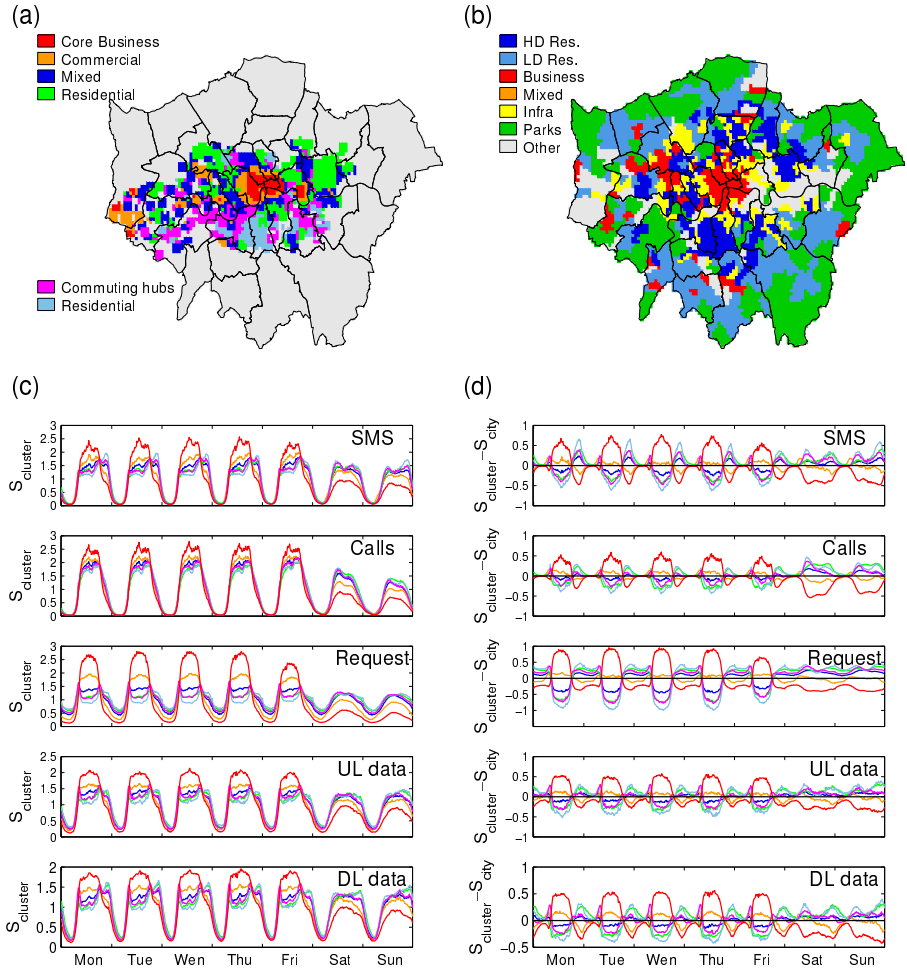}
   \end{center}
   \caption{{\small {\bf Greater London clusters}. {\bf (a)} Spatial projection of $K=6$ clusters, with their interpretation in legend (see details in main text). {\bf (b)} Actual land use maps as extracted from census data. {\bf (c)} Signatures of the clusters in the different components of activity. {\bf (d)} Deviations of the signatures compared to the whole city signatures displayed on Figure \ref{fig4}. Colors on the signatures plots match those on the cluster map (a), grey areas correspond to zones with no recorded data. \label{KAllGL}}}
\end{figure}

\begin{figure}[h!]
   \begin{center}
      \includegraphics[width=0.99\textwidth,trim = 0cm 0cm 0cm 0cm, clip]{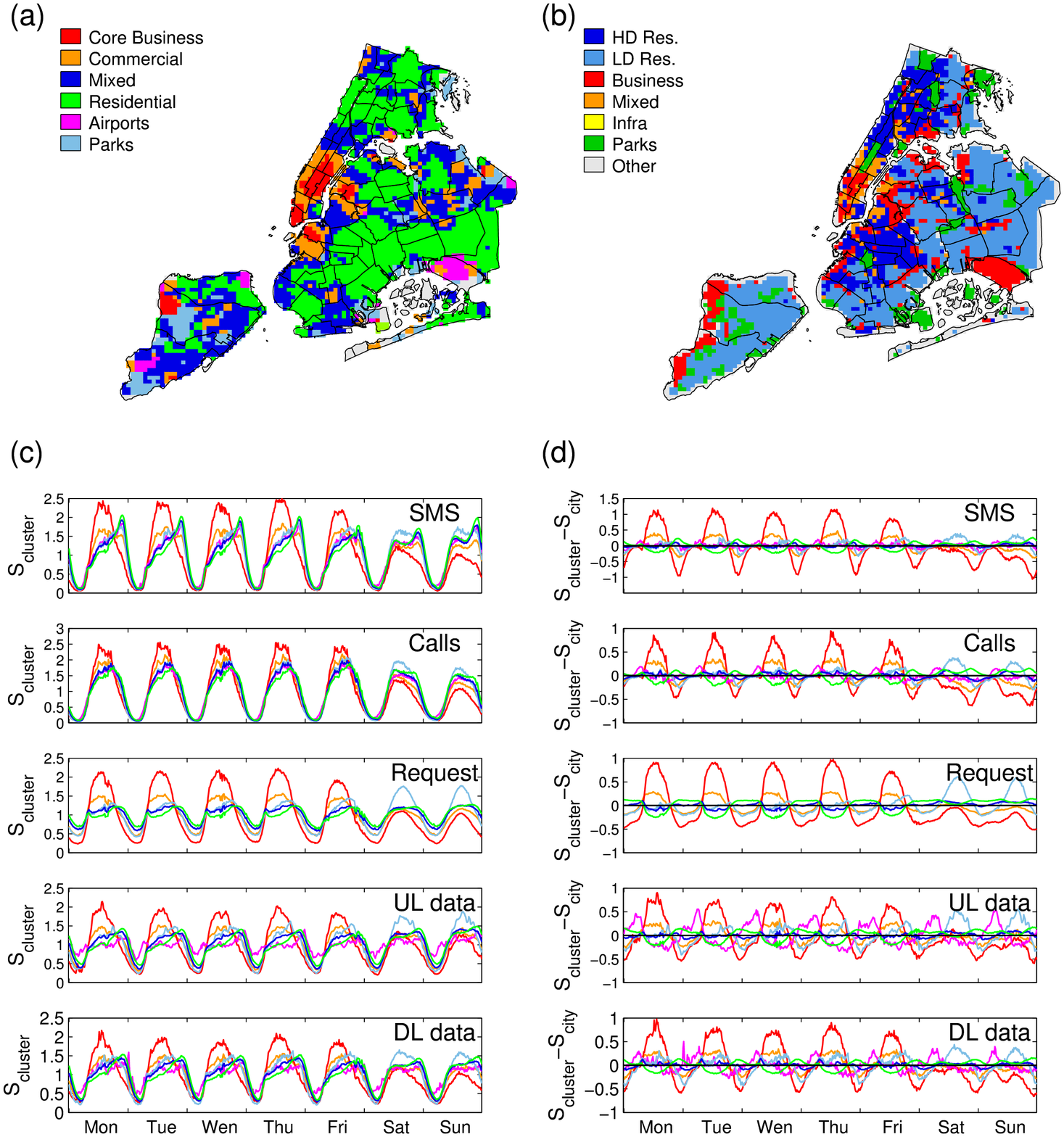}
   \end{center}
   \caption{{\small {\bf New York clusters}. {\bf (a)} Spatial projection of $K=6$ clusters, with their interpretation in legend (see details in main text). {\bf (b)} Actual land use maps as extracted from census data. {\bf (c)} Signatures of the clusters in the different components of activity. {\bf (d)} Deviations of the signatures compared to the whole city signatures displayed on Figure \ref{fig4}. Colors on the signatures plots match those on the cluster map (a), grey areas correspond to zones with no recorded data. \label{KAllNY}}}
\end{figure}

\begin{figure}[h!]
   \begin{center}
      \includegraphics[width=0.99\textwidth,trim = 0cm 0cm 0cm 0cm, clip]{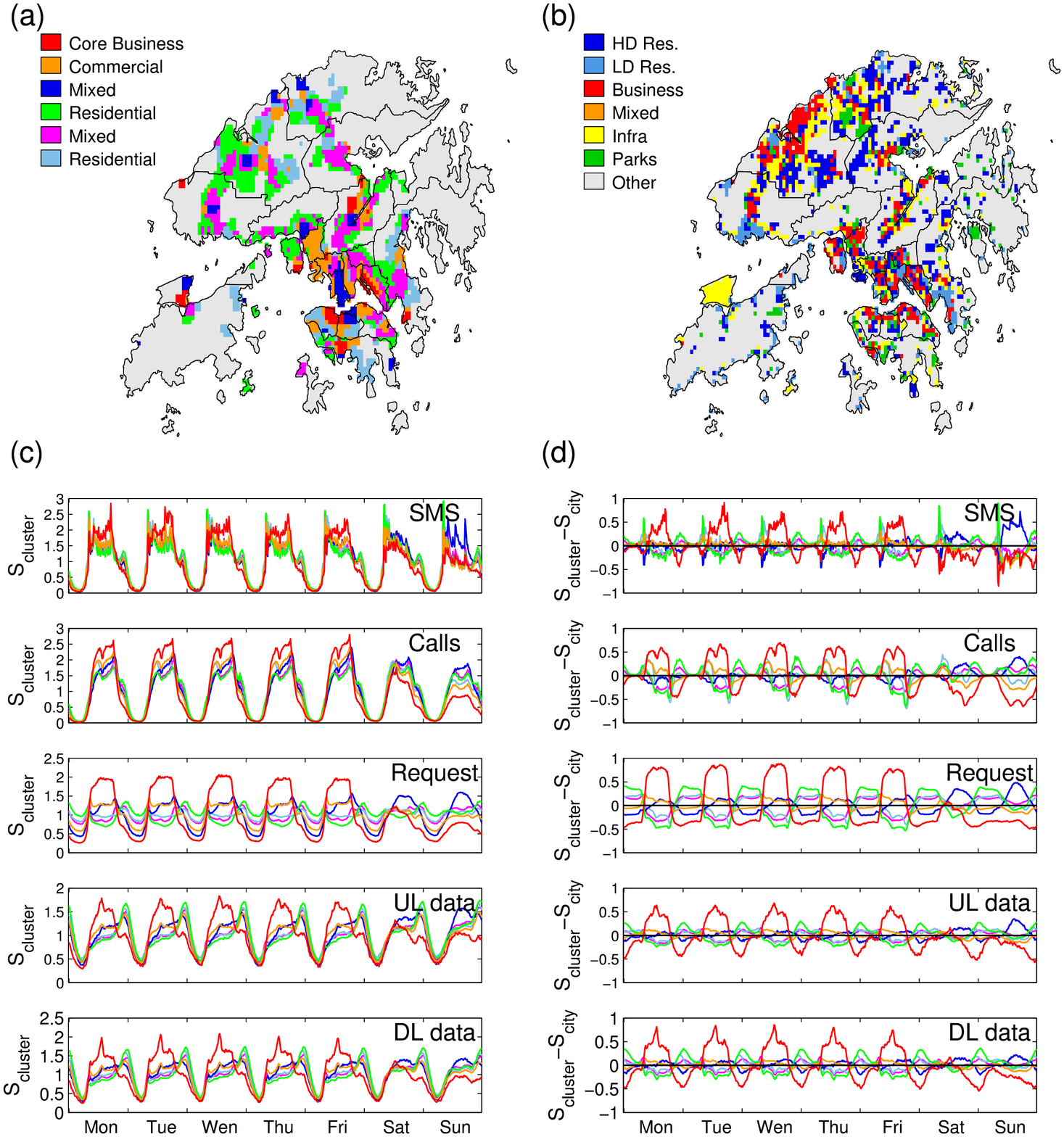}
   \end{center}
   \caption{{\small {\bf Hong Kong clusters}. {\bf (a)} Spatial projection of $K=6$ clusters, with their interpretation in legend (see details in main text). {\bf (b)} Actual land use maps as extracted from census data. {\bf (c)} Signatures of the clusters in the different components of activity. {\bf (d)} Deviations of the signatures compared to the whole city signatures displayed on Figure \ref{fig4}. Colors on the signatures plots match those on the cluster map (a), grey areas correspond to zones with no recorded data. \label{KAllHK}}}
\end{figure}

\begin{table}[hb!]
\caption{ {\bf Clusters' mobile traffic properties.} We report the relative areas covered by each cluster presented in Figures \ref{KAllGL}(a) to \ref{KAllHK}(a), as well as their share of total activities (obtained by summing the unnormalized signatures $A_i^{\lambda}$ of the cluster's pixels). \label{dataprop} }
\begin{tabular}{ l | r | rrrrr}
cluster & Area ($\%$) & DL data ($\%$) &  UL data ($\%$) &   Request ($\%$) &  Calls ($\%$)&  SMS ($\%$)  \\
\hline\noalign{\smallskip}
&&\multicolumn{4}{c}{GREATER LONDON} &\\
\hline\noalign{\smallskip}
1 (red) & 6.3 & 21.2 & 24.0 & 30.7 & 23.4 & 24.0\\
2 (orange) & 11.1 & 17.2 & 18.5 & 14.8 & 17.7 & 16.6\\
3 (blue) & 27.3 & 22.6 & 22.1 & 19.5 & 22.4 & 21.5\\
4 (green) & 23.3 & 15.6 & 13.6 & 14.4 & 15.0 & 14.7\\
5 (pink) & 18.8 & 12.2 & 11.6 & 10.3 & 11.0 & 11.4\\
6 (pastel) & 11.5 & 6.4 & 5.7 & 5.4 & 5.4 & 5.8\\
\hline\noalign{\smallskip}
&&\multicolumn{4}{c}{NEW YORK} &\\
\hline\noalign{\smallskip}
1 (red) & 3.9 & 8.7 & 11.0 & 10.9 & 8.7 & 8.1\\
2 (orange) & 10.4 & 8.9 & 10.0 & 9.7 & 8.5 & 8.3\\
3 (blue) & 32.1 & 24.7 & 24.4 & 24.1 & 24.3 & 24.8\\
4 (green) & 42.9 & 55.8 & 52.8 & 53.5 & 56.3 & 56.9\\
5 (pastel) & 7.6 & 1.2 & 1.2 & 1.2 & 1.4 & 1.3\\
6 (pink) & 2.8 & 0.6 & 0.6 & 0.5 & 0.7 & 0.6\\
\hline\noalign{\smallskip}
&&\multicolumn{4}{c}{HONG KONG} &\\
\hline\noalign{\smallskip}
1 (red) & 5.4  & 7.6 & 7.4 & 9.3 & 11.2 & 10.7\\
2 (orange) & 15.0 & 19.4 & 18.5 & 20.6 & 19.6 & 20.7\\
3 (blue) & 7.3 & 23.8 & 24.7 & 22.0 & 27.0 & 22.8\\
4 (green) & 25.8 & 12.5 & 12.2 & 12.5 & 9.8 & 10.2\\
5 (pink) & 23.0 & 25.9 & 25.4 & 26.8 & 24.0 & 22.4\\
6 (pastel) & 19.7 & 8.5 & 9.3 & 6.9 & 6.7 & 11.0\\
\hline\noalign{\smallskip}
\end{tabular}
\end{table}

\begin{table}[ht!]
\caption{{\bf Clusters' census properties.} This table shows the average population and job densities on the area covered by each cluster presented in Figures \ref{KAllGL}(a) to \ref{KAllHK}(a), interpolated from census values. Land use confusion matrices display, for each cluster, the percentage area that can be attributed to each of seven land use classes: High density residential (HD Res), Low density residential (LD Res), Business \& Commercial (Business), Mixed, Infrastructure \& Government Facilities (Infra), Parks \& Green Areas (Parks) and Other. \label{censusprop} }
\begin{tabular}{ l | rr | rrrrrrrr}
cluster & Pop$/km^2$ & Job$/km^2$ & HD Res & LD Res & Business & Mixed & Infra & Parks & Other \\
\hline\noalign{\smallskip}
& & \multicolumn{3}{c}{GREATER LONDON} &\multicolumn{4}{c }{}\\
\hline\noalign{\smallskip}
1 (red) & 8,310 & 48,390  & 0.00 & 0.00 & 73.64 & - & 1.82 & 5.45 & 19.09\\
2 (orange) & 9,450 & 9,492  & 3.59 & 1.03 & 37.44 & - & 18.46 & 5.64 & 33.85\\
3 (blue) & 9,977 & 4,084  & 18.66 & 5.66 & 14.26 & - & 14.88 & 11.53 & 35.01\\
4 (green) & 11,341 & 3,338  & 35.63 & 8.60 & 7.62 & - & 31.20 & 2.21 & 14.74\\
5 (pink) & 8,512 & 2,743 & 23.40 & 8.51 & 3.04 & - & 19.15 & 16.11 & 29.79\\
6 (pastel) & 10,257 & 2,430 & 55.94 & 16.34 & 1.49 & - & 17.82 & 1.98 & 6.44\\
\hline\noalign{\smallskip}
& & \multicolumn{3}{c}{NEW YORK} &\multicolumn{4}{c }{}\\
\hline\noalign{\smallskip}
1 (red) & 19,669 & -  & 6.84 & 3.42 & 64.10 & 14.53 & - & 11.11 & -\\
2 (orange) & 20,881 & -  & 16.03 & 28.21 & 35.26 & 10.26 & - & 10.26 & -\\
3 (blue) & 15,554 & -  & 17.22 & 51.14 & 15.56 & 3.01 & - & 13.07 & -\\
4 (green) & 17,310 & -  & 26.65 & 52.21 & 10.49 & 2.41 & - & 8.24 & -\\
5 (pink) & 3,002 & -  & 1.20 & 24.10 & 65.06 & 0.00 & - & 9.64 & -\\
6 (pastel) & 6,972 & -  & 2.62 & 43.23 & 22.27 & 0.00 & - & 31.88 & -\\
\hline\noalign{\smallskip}
& & \multicolumn{3}{c}{HONG KONG} &\multicolumn{4}{c }{}\\
\hline\noalign{\smallskip}
1 (red) & 24,722 & -  & 6.74 & 16.85 & 32.58 & - & 22.47 & 4.49 & 16.85\\
2 (orange) & 24,336 & -  & 17.41 & 8.10 & 22.67 & - & 19.03 & 7.29 & 25.51\\
3 (blue) & 27,570 & -  & 14.05 & 3.31 & 27.27 & - & 23.97 & 13.22 & 18.18\\
4 (green) & 12,121 & -  & 29.18 & 7.53 & 10.12 & - & 13.88 & 7.06 & 32.24\\
5 (pink) & 16,713 & -  & 18.68 & 11.32 & 16.58 & - & 17.63 & 5.00 & 30.79\\
6 (pastel) & 10,828 & -  & 26.77 & 6.15 & 8.31 & - & 20.62 & 4.00 & 34.15\\
\hline\noalign{\smallskip}
\end{tabular}
\end{table}

A notable drawback of the K-means algorithm is the difficulty to determine the `best' number $K$ of clusters, whose value can depends on the shape and scale of the distribution of points in a data set and the desired clustering resolution. Different ad-hoc techniques to make that decision exist, most of them based on finding the values of $K$ which balancing the search for minimizing the intra-cluster distance $E_K$ and maximizing the inter-cluster distances. There is however no consensus on the best method to use and the correct choice of $K$ may also often rely on the researchers' expert opinion and search for interpretable results.
We guided our choice by looking at local maxima of the silhouette index \cite{rousseeuw1987silhouettes}. All cities presented local maxima for $K=2$ clusters (corresponding roughly to city centers and city suburbs), $K=6$ and larger values of $K$ which vary with the studied city. All results presented in the following have been obtained for $K=6$, which is the most relevant case. We indeed found that allowing a larger number of clusters mostly added clusters concentrated of very few pixels in areas with very low mobile phone activity and without any regular signatures.

%----------------------------------------------------------------------------------------------------------------------------------------------------------------------------
\subsection{Revealing the spatial structures of cities}
\label{section::cluster-city}
We conducted an independent K-means clustering analysis for each city. As we previously stated, the best cluster size distribution and interpretability was achieved for $K=6$ in each case.
Figures \ref{KAllGL}(a) to \ref{KAllHK}(a) show the spatial projections of the clusters on a map of the cities. First of all, it is worth noting that the clusters are made of spatially cohesive groups of pixels shaping concentric-like structure within the cities. The signatures of the clusters as well as their deviation from their city signature are displayed on Figures \ref{KAllGL}(c-d) to \ref{KAllHK}(c-d), and Table \ref{dataprop} lists the share of total activities occurring within the surface covered by the clusters. To better understand the nature of the clusters and their relation with standard land use classification, Figures \ref{KAllGL}(b) to \ref{KAllHK}(b) display land use maps of the cities built from extracted census data (see section \ref{section::data})). Finally, Table \ref{censusprop} lists average population and job densities within the clusters (when available), as well as confusion matrices highlighting the similarities between our clusters and the land use classes.   

A few patterns are easy to interpret and compare with census data. For example:
\begin{itemize}
\item Clusters 1 (in red). In all cities, these clusters's signatures present high levels of activities during working hours and very low levels of activity in the evening during the workdays, and a huge weekday-to-weekend activity ratios. The combination of their small area shares ($4$ to $6\%$ of the cities' area as reported in Table \ref{dataprop}) and their large activity share is consistent with a high concentration of activity and an identification as business cores. This is verified since the red clusters cover the City district in London, the financial and decisional districts in New York (for example south Manhattan where Wall Street is located) and a large part of the Central district on Hong Kong Island. The confusion matrices of Table \ref{censusprop} confirm that the areas covered by clusters 1 correspond to Business \& Commercial land use.
\item Clusters 2 (in orange). Compared to the average cities signatures, orange clusters are characterized by higher levels of activity during weekday'  working hours and lower activity level in the week end, similar (but to a lower extent) than what happens in clusters 1. On London and New York maps, these clusters mostly surround the red ones (notice also that in London the orange cluster covers Heathrow Airport on the West side of Greater London). Based on these characteristics orange clusters could be identified as commercial areas. This is confirmed by the confusion matrices, reporting low to average residential land use and higher than average Business land use in these clusters. 
\item Clusters 4 (in green). Compared to the average cities signatures, these clusters display higher activity level at night and lower activity level on monday to friday business hours. The curves showing the deviations of their signatures compared to the cities signatures remarkably mirror those corresponding to clusters 1 core business areas. These features suggests that these clusters correspond to purely residential areas, which is confirmed by by the confusion matrices reporting high residential land use.
\item Clusters 3 (in blue). These clusters have signatures similar to those of residential clusters 4, but with smaller deviations from the corresponding city signatures. These features suggest, as baked up by land use data,  to interpret these clusters as mixed area with strong residential component.
\end{itemize}

Other clusters have a more city specific interpretation:
\begin{itemize}
\item Clusters 5 (in pink). These clusters have different interpretations in each city. In London, the pink cluster signatures are close to the green ones (residential), but the request and data signatures present specific peaks at commuting hours, consistent with the commuting hub nature of the area. In New York, the pink cluster mostly corresponds to the JFK airport, characterized by specific bumps of activity in the morning. In Hong Kong, these match a mixed area with a strong residential component. 
\item Clusters 6 (in pastel). In Greater London, the pastel cluster is another mainly residential cluster, in the South part of Inner London. Its signatures' properties are very similar to those of the London green cluster, with exception of a specific peak of activity on saturday just before lunch time. This peak could be explained by a recurring event (such as a market). In New York, a quick comparison with a detailed maps of the city reveals that the pastel cluster correspond to parks. The activity shares of this cluster is very low (in accordance with the fact that there are no people residing or working full time on the park premises ), and its signature show high activity levels during the weekends (when people may go for a walk within the nearer park). Finally in Hong Kong, the pink cluster appears to be a second residential one. 
\end{itemize}

To summarize, each city can be characterized by a gradation of clusters. A feature common to all cities is the existence of core business areas and purely residential areas, an already known urban fact that we were able to check thanks to communication traffic data. 

Compared to classical time consuming and expensive field surveys, our approach make it possible to built automatic, quick and relatively cheap way of preparing land use maps of the cities. The maps we obtained are closely related to classic land use maps, especially for the distinction between Business and residential areas. When it comes to other land use classes, some difference occurs. Indeed, while the classical land use classes have to with what the land looks like (is the neighborhood consisting of retail store, bank, residential buildings, etc), the clusters we found are based on communication data revealing human dynamic behaviors (like working in an office, shopping, eating, commuting, sleeping, etc). Rather than emulating classical land use mapping, our approach thus produces a complementary point of view that enrich our understanding of the multiple dynamics at stake in the cities. 

Although we used every type of activity in our clustering analyses, the similarities between the signatures of different type (Calls, SMS, Request, UL or DL Data) displayed on Figures \ref{KAllGL}(c) to \ref{KAllHK}(c) suggests that our finding would not qualitatively change if we focused on only one activity type.

%----------------------------------------------------------------------------------------------------------------------------------------------------------------------------
\subsection{Revealing Universal Patterns  }
\label{section::cluster cities}

\begin{figure}[b!]
   \begin{center}
      \includegraphics[width=1\textwidth]{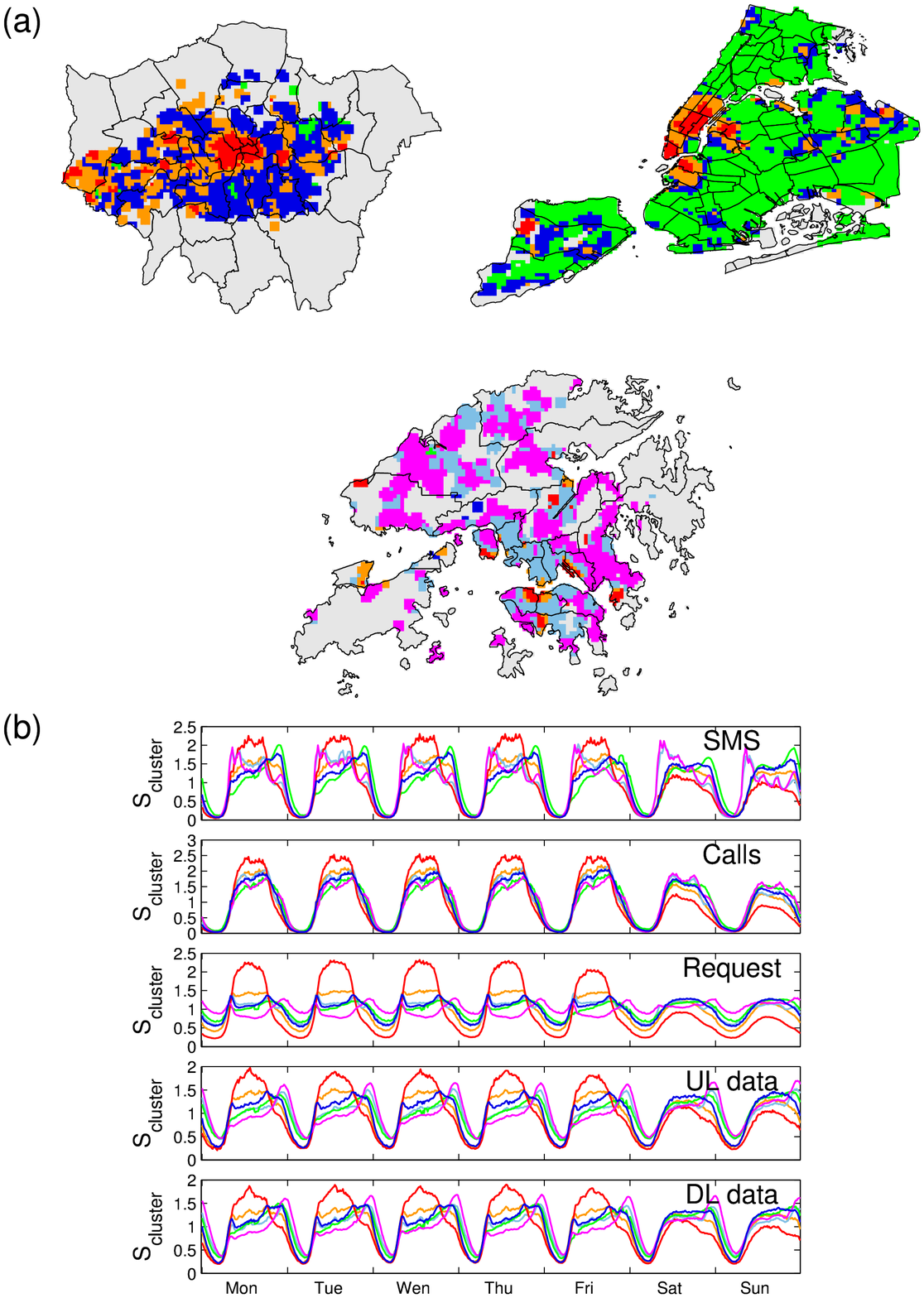}
     \end{center}
        \caption{{\small {\bf Clustering the three cities at once}. {\bf (a)} Spatial projection of $K=6$ clusters of all cities' counter data (grey areas correspond to zones with no recorded data). {\bf (b)} Clusters' Signatures. Colors on the plots match those on the map. \label{Kallall}}}
\end{figure}

The results presented in the previous section suggest that mobile traffic patterns can reveal concentric structure of cities into clusters that can be interpreted the same way. To what extent are these cities similar? In this section, we investigate this issue by making a transversal analysis of our three studied cities. We performed a K-means clustering analysis on all cities at once, grouping 500m by 500m grid pixels with similar signature patterns.  

Figure \ref{Kallall} displays the results of this transversal clustering analysis. As before, we chose to display results obtained for $K=6$.

\begin{itemize}
\item All previously identified core business centers are gathered into a single (here red) cluster, whose signature can be characterized as before. 
\item Similarly, the previously identified `commercial' areas are roughly gathered into a single cluster - here orange. The correspondence between this orange cluster and those found in city-independent analyses appear is evident for New York, but less obvious in London and Hong Kong (see Figures \ref{KAllGL}(a) to \ref{KAllHK}(a)). 
\item The other clusters mostly correspond to the previously identified residential areas. Surprisingly, these clusters are almost completely concentrated in one city: the blue cluster is specific to London, the green cluster is specific to New York and the pink and pastel clusters are specific to Hong kong.   
\end{itemize}

Concerning the residential area, the transversal clustering analysis emphasize the differences due to local cultural, technological and economical factors identified in section \ref{section::typweekcities}, e.g. the evening peaks of SMS in New York, or the evening peaks of Data transfer in Hong Kong. The very strong and somewhat surprising result here is the fact that the studied cities have core business centers that share a similar pattern despite those local factors.

%----------------------------------------------------------------------------------------------------------------------------------------------------------------------------
%----------------------------------------------------------------------------------------------------------------------------------------------------------------------------

\section{Discussion}
 \label{section::discussion} 
Our research findings demonstrate that a general understanding of the mobile network signatures can help us to look at cities with a renewed perspective.

We saw in section \ref{section::spatial} how time-aggregated maps of mobile traffic inhomogeneities could capture spatial patterns revealing locations where people are in general most active. This approach allows to track the location where people spend most of their time and is complementary to more traditional census data recording where people live or work. Doing a similar analysis on specific periods of time, one can expect a rather good correlation between mobile phone activity and job density, especially during working hours, and a relatively less good correlation with residential density during working hours that would increase when people are typically at home (late in the evenings, early in the mornings or during the week ends).

In addition, these maps are more up-to-date than those based on census polling, relatively cheaper to obtain and dynamic. From a research point of view, one could also imagine to use insights gain from such representation (like the Gini indices measuring spatial inhomogeneities) to enrich current taxonomies of cities.

In section \ref{section::temporal}, we defined typical week patterns, or {\em signatures} to characterize the activities dynamics at city or local scale. By comparing city signatures, we highlighted specific influencing factors (mobile traffic plan policy, technological, economical and cultural factors) shaping those dynamics in Greater London, New York and Hong Kong. Building on the example of a few selected locations within London, we showed how the signatures could reveal the nature (either financial, commercial, recreational, residential, commuting hub, etc.) of the concerned areas. 
In general, the insights gained from the study of the typical week signatures could be used to optimize the overall network performance by informing the mobile operators of the actual typical usage for them to take proactive decisions upon. Diffusing the knowledge of the signature patterns could also generate new business ideas in the cities, by just allowing retailers to optimize scheduling based on the likelihood of people being in the vicinity of their stores.

We presented in section \ref{section::cluster} a clustering analysis process allowing automatic detection of locations with similar signatures within a city. Our findings showed that we could detect in each city a core business center, at least one pure residential area, one mixed area with a strong commercial component and another mixed area with a strong residential component. In addition, we were able to detect other clusters of specific nature (commuting hub areas in London or the JFK airport in New York). The methods used could be easily pushed to generalize these findings. 
We stressed that our approach make it possible to built automatic, quick and relatively cheap way of preparing maps that could complement and enrich classical land use maps based on surveys. Indeed, communication traffic data tell us about the actual dynamical behavior of people at each location, while land use maps rather tell us about the average type of behavior you can expect based on the urbanization levels and the type of buildings, shops or infrastructures present at each location. 

Our final finding was obtained by applying our clustering procedure on the locations of all three cities at once. Quite surprisingly, we found that the core business centers of London, New York and Hong Kong were gathered into a single cluster, which prove their high degree of similitude. On the other hand, the city have residential locations whose signatures are well distinct. 
To answer a question raised at the beginning of this chapter, it seems like globalization shapes the economical and political activity in large cities' financial and decisional core centers, while individuals activity patterns are still defined mostly by local factors.

On a broader note, as the digital databases are growing and our computational methods are improving, it may be tempting to multiply automatic procedures to generate lists of insights based on mobile traffic data. However, we argue that knowledge expertise is more than ever needed to understand, interpret and critic these results. 

Though this chapter presented new similarity measurements and links between three major cities, it also opens up the question of the universality of our findings. Is the common beat detected in the core areas a universal pattern common to all cities, or is it only peculiar to `occidental' developed cities? Is there any natural classification of the world's cities based on similarity between peripheral residential signatures? It would be most interesting to study these questions by enlarging our datasets to include major cities from both developed and developing countries such as Paris, Mexico City, Shanghai, Rio de Janeiro, Sao Paulo, Lagos, or Mumbai.
The challenge is now to gather a collection of mobile traffic data from all these cities before starting to built a yet-to-define comparative science of cities based on them.

%----------------------------------------------------------------------------------------------------------------------------------------------------------------------------
%----------------------------------------------------------------------------------------------------------------------------------------------------------------------------

%
\subsection*{acknowledgement}
We thank Ericsson for providing datasets for this research, and especially Dwight Witherspoon for organizational support to the project. We also thank Christine Mayni\'e-Fran\c cois for stimulating discussions and thorough proofreading.
We would further like to thank the National Science Foundation, the AT\&T Foundation, the Rockefeller Foundation, the MIT SMART program, the MIT CCES program, Audi Volkswagen, BBVA, The Coca Cola Company, Ericsson, Expo 2015, Ferrovial, the Regional Municipality of Wood Buffalo and all the members of the MIT Senseable City Lab Consortium for supporting this research.

\end{document}